\documentclass[aps,prb,preprint,superscriptaddress,showpacs]{revtex4-1}

\usepackage{graphicx}
\usepackage[usenames]{color}
\usepackage{amssymb}
\usepackage{amsmath}
\begin{document}


\title{Thickness-dependent ferromagnetic metal to paramagnetic insulator 
transition in La$_{0.6}$Sr$_{0.4}$MnO$_3$ thin films studied by 
x-ray magnetic circular dichroism}


\author{G. Shibata}
\email[]{shibata@wyvern.phys.s.u-tokyo.ac.jp}
\affiliation{Department of Physics, University of Tokyo, Bunkyo-ku, Tokyo 
113-0033, Japan}

\author{K. Yoshimatsu}
\affiliation{Department of Physics, University of Tokyo, Bunkyo-ku, Tokyo 
113-0033, Japan}
\affiliation{Department of Applied Chemistry, University of Tokyo, Bunkyo-ku, 
Tokyo 113-8656, Japan}

\author{E. Sakai}
\affiliation{Department of Applied Chemistry, University of Tokyo, Bunkyo-ku, Tokyo 
113-8656, Japan}
\affiliation{Photon Factory, Institute of Materials Structure Science, High 
Energy Accelerator Research Organization (KEK), Tsukuba, Ibaraki 305-0801, 
Japan}

\author{V. R. Singh}
\affiliation{Department of Physics, University of Tokyo, Bunkyo-ku, Tokyo 
113-0033, Japan}

\author{V. K. Verma}
\affiliation{Department of Physics, University of Tokyo, Bunkyo-ku, Tokyo 
113-0033, Japan}

\author{K. Ishigami}
\affiliation{Department of Complexity Science and Engineering, 
University of Tokyo, Bunkyo-ku, Tokyo 113-0033, Japan}

\author{T. Harano}
\affiliation{Department of Physics, University of Tokyo, Bunkyo-ku, Tokyo 
113-0033, Japan}

\author{T. Kadono}
\affiliation{Department of Physics, University of Tokyo, Bunkyo-ku, Tokyo 
113-0033, Japan}

\author{Y. Takeda}
\affiliation{Condensed Matter Science Division, Japan Atomic Energy Agency, 
Sayo-cho, Sayo-gun, Hyogo 679-5148, Japan}

\author{T. Okane}
\affiliation{Condensed Matter Science Division, Japan Atomic Energy Agency, 
Sayo-cho, Sayo-gun, Hyogo 679-5148, Japan}

\author{Y. Saitoh}
\affiliation{Condensed Matter Science Division, Japan Atomic Energy Agency, 
Sayo-cho, Sayo-gun, Hyogo 679-5148, Japan}

\author{H. Yamagami}
\affiliation{Condensed Matter Science Division, Japan Atomic Energy Agency, 
Sayo-cho, Sayo-gun, Hyogo 679-5148, Japan}
\affiliation{Department of Physics, Kyoto Sangyo 
University, Kyoto 603-8555, Japan}

\author{A. Sawa}
\affiliation{National Institute of Advanced Industrial Science and Technology 
(AIST), Tsukuba, Ibaraki 305-8562, Japan}

\author{H. Kumigashira}
\affiliation{Department of Applied Chemistry, University of Tokyo, Bunkyo-ku, Tokyo 
113-8656, Japan}
\affiliation{Photon Factory, Institute of Materials Structure Science, High 
Energy Accelerator Research Organization (KEK), Tsukuba, Ibaraki 305-0801, 
Japan}

\author{M. Oshima}
\affiliation{Department of Applied Chemistry, University of Tokyo, Bunkyo-ku, Tokyo 
113-8656, Japan}

\author{T. Koide}
\affiliation{Photon Factory, Institute of Materials Structure Science, High 
Energy Accelerator Research Organization (KEK), Tsukuba, Ibaraki 305-0801, 
Japan}

\author{A. Fujimori}
\affiliation{Department of Physics, University of Tokyo, Bunkyo-ku, Tokyo 
113-0033, Japan}
\affiliation{Department of Complexity Science and Engineering, University of 
Tokyo, Bunkyo-ku, Tokyo 113-0033, Japan}
\affiliation{Condensed Matter Science Division, Japan Atomic Energy Agency, 
Sayo-cho, Sayo-gun, Hyogo 679-5148, Japan}


\date{\today}

\begin{abstract}
Metallic transition-metal oxides undergo a metal-to-insulator 
transition (MIT) as the film thickness decreases across a critical thickness 
of several monolayers (MLs), but its driving mechanism 
remains controversial. 
We have studied the thickness-dependent MIT of the ferromagnetic metal 
La$_{0.6}$Sr$_{0.4}$MnO$_3$ 
by x-ray absorption spectroscopy and x-ray magnetic circular dichroism. 
As the film thickness was decreased across the critical thickness 
of the MIT (6-8 ML), a gradual decrease of the ferromagnetic signals and 
a concomitant increase of paramagnetic signals were observed, 
while the Mn valence abruptly decreased towards Mn$^{3+}$. 
These observations suggest that the ferromagnetic phase gradually and 
most likely inhomogeneously turns into the paramagnetic phase 
and both phases abruptly become insulating at the critical thickness. 
\end{abstract}


\maketitle


\section{\label{Intro}Introduction}
The physical properties of 3$d$ transition-metal oxides (TMOs) 
are usually controlled by the bandwidth and/or 
band filling of the 3$d$ bands \cite{MITrev}. 
Recently, attempts have also been made to control them by the 
thickness of thin film samples, namely, by dimensionality 
\cite{HongLSMO, HuijbenLSMO, YoshiLSMO, LSMO110_PRL12, XiaSRO, 
ToyotaSRO, YoshiSVO10,YoshiSVO11, LNOthickdep}. 
In many metallic oxide thin films, including ferromagnetic ones such as 
La$_{1-x}$Sr$_{x}$MnO$_3$ (LSMO) 
\cite{HongLSMO, HuijbenLSMO, YoshiLSMO, LSMO110_PRL12} and 
SrRuO$_3$ (SRO) \cite{XiaSRO, ToyotaSRO}, 
and paramagnetic ones such as SrVO$_3$ (SVO) 
\cite{YoshiSVO10,YoshiSVO11} and LaNiO${_3}$ \cite{LNOthickdep}, 
the resistivity increases when the film thickness is decreased, 
and metal-to-insulator transitions (MITs) occur 
at a critical thickness of several monolayers (MLs). 
Evidence for the MITs was found by transport measurements 
\cite{HongLSMO, HuijbenLSMO, XiaSRO, LNOthickdep}
and photoemission spectroscopy (PES) 
\cite{YoshiLSMO, ToyotaSRO, YoshiSVO10, YoshiSVO11}. 
According to the PES studies of TMO thin films, 
the density of states at the Fermi level ($E_F$) disappears below 
4 ML (SRO \cite{ToyotaSRO}, SVO \cite{YoshiSVO10}) to 8 ML 
(LSMO \cite{YoshiLSMO}), 
resulting in a large insulating gap (of order $\sim$ 1 eV) at $E_F$. 
In order to explain such MITs, transport theories 
for conventional metal thin films \cite{FSeq1, FSeq2}, 
which take into account only surface roughness, 
predict extremely small critical thicknesses, 
and one has to invoke strong electron-electron scattering, 
namely, strong electron correlation. 
For example, the thickness-dependent MIT of SVO thin films has been considered 
as a bandwidth-controlled Mott transition caused by the decreased number of 
nearest-neighbor V atoms \cite{YoshiSVO10}. 
Furthermore, the decrease of the film thickness leads to 
the lowering of spatial dimension and symmetry, 
and the increase of interfacial effects, 
resulting in the changes of the electric and magnetic properties. 
In LSMO \cite{HuijbenLSMO, YoshiLSMO} as well as in SRO \cite{XiaSRO}, 
not only metallic conduction but also ferromagnetism 
disappears simultaneously. 
For LSMO thin films, where the number of electrons is not an integer, 
the simple bandwidth-controlled Mott transition mechanism alone 
is not sufficient to explain the MIT and one has also to take into account 
the effect of disorder or local lattice distortion to induce the localization 
of charge carriers. 
Therefore, it is important to consider mechanisms such as 
charge ordering and/or the splitting of the $d$ bands due to the lower 
structural symmetry at the surface and interface. 
Such mechanisms are usually accompanied by changes in the magnetic 
properties. 
Thus, it is strongly desired to probe both the electronic states 
and magnetic properties on the microscopic level at the same time 
as a function of film thickness. 

For this purpose, we have investigated the electric and magnetic properties 
of LSMO ($x=0.4$) thin films as functions of film thickness 
using x-ray magnetic circular dichroism (XMCD). 
XMCD in core-level x-ray absorption spectroscopy (XAS) is a powerful tool 
to obtain information about the magnetism of specific elements, 
together with information about the valence states, and is especially suitable 
for the study of thin films and nanostructures, because it can probe the 
intrinsic magnetism without contribution from the substrate 
and other extrinsic effects. 

\section{\label{Exp}Experiment}
LSMO thin films with various thicknesses were fabricated on the 
TiO$_2$-terminated (001) surface of SrTiO$_3$ (STO) substrates 
by the laser molecular beam epitaxy (laser-MBE) method. 
The LSMO thickness ranged from 2 ML to 15 ML. 
After the deposition, the films were capped with 1 ML of 
La$_{0.6}$Sr$_{0.4}$TiO$_3$ (LSTO) and then 
2 ML of STO [Fig.\ \ref{XASall}(a)]. 
The LSTO layer was inserted to keep the local environment of the topmost 
MnO$_2$ layer the same as that of the deeper MnO$_2$ layers \cite{YoshiLSMO}, 
namely, each MnO$_2$ layer is sandwiched by 
La$_{0.6}$Sr$_{0.4}$O (LSO) layers. 
All the fabrication conditions were identical to those of 
Ref.\ \onlinecite{YoshiLSMO}. 
The surface morphology of the multilayers was checked by 
atomic force microscopy, and atomically flat step-and-terrace structures 
were clearly observed in all the films studied. 
Four-circle x-ray diffraction measurements confirmed the 
coherent growth of the films. 
Prior to measurements, the samples were annealed in 1 atm of O$_2$ 
at 400 $^{\circ}\text{C}$ for 45 minutes 
in order to eliminate oxygen vacancies. 
The XAS and XMCD measurements were performed at 
polarization-variable undulator beamlines BL-16A of the Photon Factory (PF) 
and at BL23SU of SPring-8 \cite{BL23SU}. 
The maximum magnetic field was 
$\mu_{0} H_{\text{ext}} = 3\ \text{T}$ at PF and 
$\mu_{0} H_{\text{ext}} = 8\ \text{T}$ at SPring-8. 
At both beamlines, the magnetic field was applied perpendicular 
to the film surface. 
Photons were incident normal to the sample surface and 
their helicity was reversed to measure XMCD. 
The measurements were performed at $T=20$ K. 
All the spectra were taken in the total electron yield mode. 
Most of the spectra shown in the figures were taken at PF, 
while data taken at high magnetic fields ($\mu_{0} H_{\text{ext}} > 3\ \text{T}$) 
were measured at SPring-8. 

\section{\label{Res}Results and Discussion}
\begin{figure}[tbp]
\centering
\includegraphics[width=8cm]{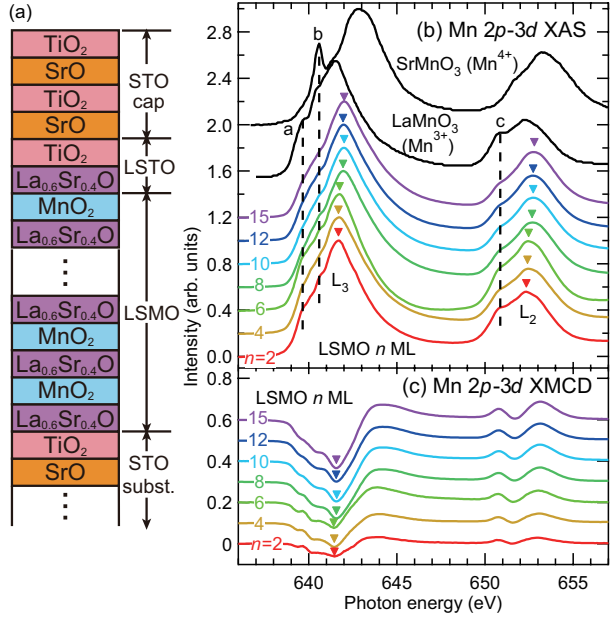}
\caption{(Color online) 
XAS and XMCD spectra of the La$_{0.6}$Sr$_{0.4}$MnO$_3$ (LSMO) thin films 
for various thicknesses. 
(a) Schematic drawing of the thin film samples consisting of LSMO, SrTiO$_3$ 
(STO), and La$_{0.6}$Sr$_{0.4}$TiO$_3$ (LSTO) layers. 
(b) Mn $2p$ XAS spectra. 
(c) Mn $2p$ XMCD spectra. 
All the spectra were taken at $T=20$ K with an external magnetic field 
$\mu_{0} H_\text{ext} = 3$ T. In panel (b), 
Mn $2p$ XAS spectra of LaMnO$_{3}$ 
(Mn$^{3+}$) \cite{Mn23XAS} and SrMnO$_{3}$ (Mn$^{4+}$) \cite{SMOXAS} 
are also shown as references. }
\label{XASall}
\end{figure}

\begin{figure}[tbp]
\centering
\includegraphics[width=8cm]{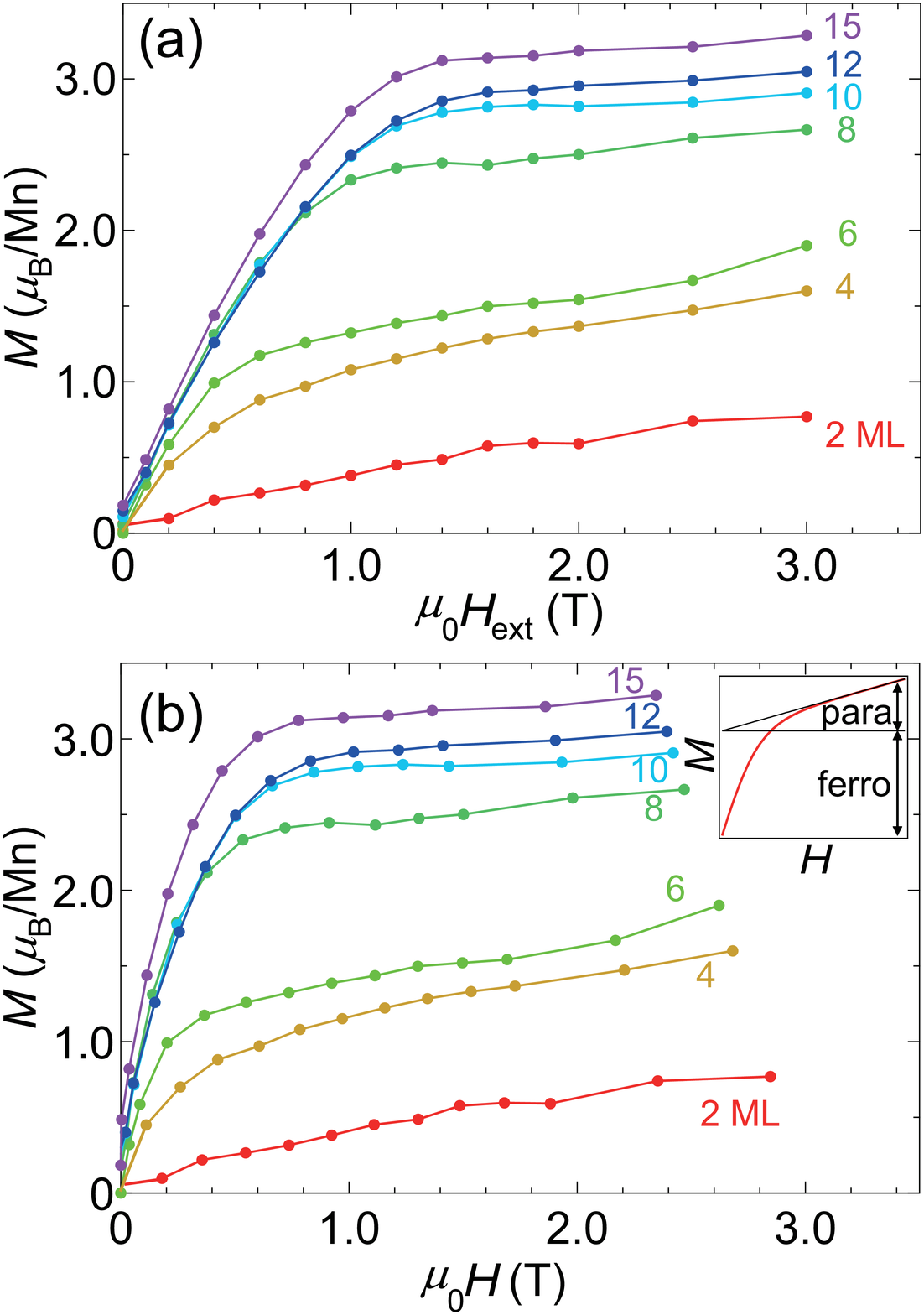}
\caption{(Color online) Magnetic field dependence of the magnetization ($M$) 
of LSMO thin films estimated by XMCD. 
(a) $M$ plotted against the external field $H_{\text{ext}}$. 
(b) $M$ plotted against the magnetic field corrected 
for the demagnetizing field (see text). 
Inset shows the decomposition of $M$ into the ferromagnetic 
($M_{\text{ferro}}$) and paramagnetic ($M_{\text{para}}$) components. 
}
\label{MHcurves}
\end{figure}

Figures \ref{XASall}(b) and \ref{XASall}(b)(c) show the XAS and XMCD spectra 
of all the LSMO samples. 
The spectral line shapes of XMCD [Fig.\ \ref{XASall}(c)] are similar to 
those of bulk LSMO \cite{KoideLSMO}. 
Reflecting the thickness-dependent magnetic properties of 
LSMO thin films \cite{HongLSMO, HuijbenLSMO, YoshiLSMO}, 
the XMCD intensity decreased with decreasing thickness. 
Using the XMCD sum rules \cite{orbsum, spinsum}, we have estimated the 
spin ($M_{\text{spin}}$) and orbital ($M_{\text{orb}}$) magnetic moments 
of the Mn ions 
\cite{[{Since the spin sum rule tends to underestimate the magnetic moment 
when spin-orbit splitting of the transition-metal $2p$ core levels is not 
large enough compared to the $2p$-$3d$ exchange splitting interaction, 
we have divided $M_{\text{spin}}$ by a 
`correction factor' 0.587 given in }][{}]Teramura}. 
The ratio $|M_{\text{orb}}/M_{\text{spin}}|$ was found to be less than 
$\sim 1/100$ for all the LSMO thicknesses; therefore
$M_{\text{spin}}$ mainly contributes to the magnetic moment of Mn. 
Figure \ref{MHcurves} shows thus estimated magnetization curves 
($M \equiv M_\text{spin} + M_\text{orb} \sim M_\text{spin}$) of the 
LSMO thin films with various thicknesses. 
In Fig.\ \ref{MHcurves}(a), $M$'s are plotted as functions of 
applied magnetic field $H_\text{ext}$. 
In Fig.\ \ref{MHcurves}(b) the same data are replotted as functions of 
magnetic field corrected for the demagnetizing field $H_\text{demag}$ 
perpendicular to the film as 
$H = H_{\text{ext}}+H_\text{demag} = H_{\text{ext}}-M$. 
After the demagnetization-field correction, the saturation field is reduced 
to $\mu_{0} H \sim 0.3$ - $0.8\ \text{T}$, 
but is still an order of magnitude larger than the saturation field 
when the external magnetic field is applied parallel to the film and 
the demagnetizing field is absent 
($\mu_{0} H \sim 10^{-2}\ \text{T}$) \cite{Ishii_APL05}. 
Therefore, we conclude that the LSMO thin films have in-plane 
easy magnetization axes due to magnetocrystalline anisotropy, 
which  probably originates from 
the tensile strain from the STO substrate 
\cite{Konishi_JPSJ, LSMOLD_Natcom12} 
(note that $a = 0.387\ \text{nm}$ for bulk LSMO while 
$a = 0.3905\ \text{nm}$ for STO). 

\begin{figure}[tbp]
\centering
\includegraphics[width=8cm]{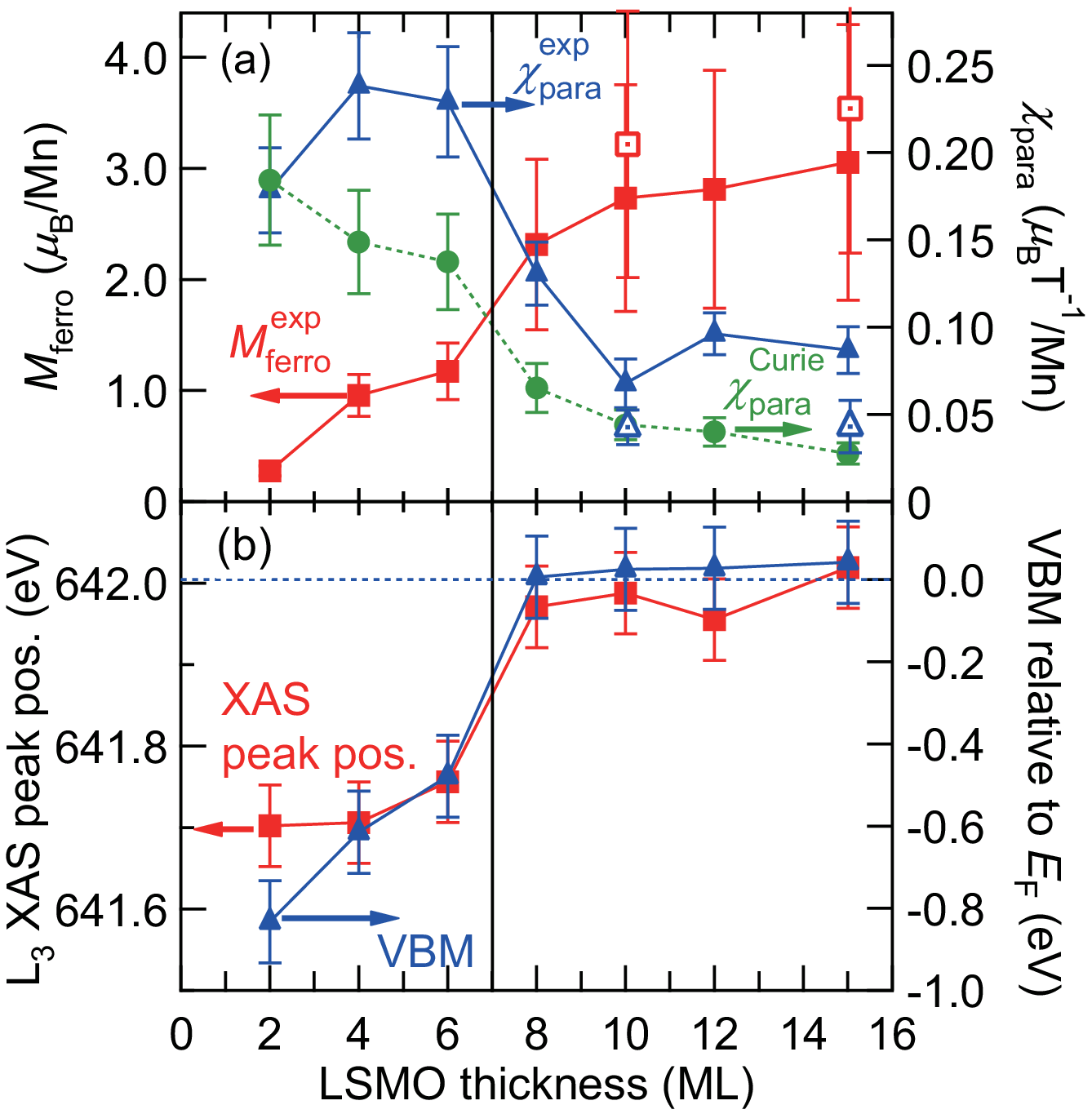}
\caption{(Color online) Thickness dependencies of the 
magnetic and electronic properties of the LSMO thin films. 
(a) Thickness dependence of the ferromagnetic moment per Mn 
($M_{\text{ferro}}$) and the paramagnetic susceptibility 
($\chi_{\text{para}}$) per Mn of the LSMO thin films, 
estimated from the magnetization curves in Fig.\ \ref{MHcurves}. 
The paramagnetic susceptibility simulated using the Curie law 
($\chi_{\text{para}}^{\text{Curie}}$) is shown by a dashed curve, 
indicating an enhancement of the paramagnetic signals. The 
$M_{\text{ferro}}$ and $\chi_{\text{para}}$ estimated from the magnetization 
measurements up to $\mu_{0} H_{\text{ext}} \leq 8\ \text{T}$ are shown by 
open symbols. 
(b) Peak position of the Mn L$_3$ edge estimated from the XAS spectra and 
the valence band maximum (VBM) position relative to the Fermi level ($E_{F}$) 
measured by PES \cite{YoshiLSMO}. 
}
\label{Ndep}
\end{figure}

Next, we decompose the magnetization curves into two components as shown 
in the inset of Fig.\ \ref{MHcurves}(b): the ferromagnetic component 
which saturates below $\mu_{0} H \sim 1\ \text{T}$ 
and the paramagnetic component which increases linearly with $H$ 
up to the highest magnetic fields
\cite{[{Examples of such a decomposition of the magnetization curves 
can be seen in references, e.g.\ }][{}]Takeda_GaMnAs}. 
Thus the intercept of the magnetization curve gives the 
ferromagnetic moment $M_{\text{ferro}}$ 
and the slope of the magnetization curve gives paramagnetic susceptibility 
$\chi_{\text{para}}$ of the Mn ions. 
We confirmed the linear increase of $M$ as a function of $H$ up to 
higher magnetic fields ($3\ \text{T} < \mu_{0} H_{\text{ext}} \leq 8\ \text{T}$) 
for several samples (not shown), which justifies the separation of the 
magnetization curves into the ferromagnetic and paramagnetic components. 
In Fig.\ \ref{Ndep}(a), the $M_{\text{ferro}}$ and $\chi_{\text{para}}$ 
values thus obtained are plotted as functions of film thickness. 
With decreasing film thickness, $M_{\text{ferro}}$ decreases and 
$\chi_{\text{para}}$ increases, indicating a gradual transition 
from the ferromagnetic state to the paramagnetic state. 
The measured $\chi_{\text{para}}$ is, however, somewhat larger than 
the one predicted by the Curie law of the 
Mn$^{3+}$-Mn$^{4+}$ mixed valence state 
in the entire thickness range
\footnote{
We have compared the magnitude of $\chi_{\text{para}}$ 
with that of non-interacting Mn$^{3+}$/Mn$^{4+}$ local moments calculated 
for the Curie paramagnetic state, namely, 
\begin{align*}
\chi_{\text{para}}^{\text{Curie}} & = \frac{S(S+1)g^2\mu_B}{3k_BT}\left( 1 - \frac{M_{\text{ferro}}}{M_0} \right),
\end{align*}
where $g \simeq 2.0$ is the $g$-factor, $M_0 = 3.6 \mu_B/\text{Mn}$ 
is the saturation magnetization of LSMO, and 
$1 - M_{\text{ferro}}/M_0$ is the number ratio of the paramagnetic Mn atoms. 
}. 
This indicates that the system exhibits a ferromagnetic-to-paramagnetic 
phase separation and that even in the paramagnetic state 
there are ferromagnetic correlations between the Mn local moments. 

We note that while the thickness-dependent MIT occurs rather abruptly 
according to PES \cite{YoshiLSMO}, 
Fig.\ \ref{Ndep}(a) shows a \textit{gradual} decrease and increase of the 
ferromagnetic and paramagnetic components, respectively, 
with decreasing film thickness. 
Indeed, the thickness dependence of the valence-band maximum (VBM) 
measured by PES \cite{YoshiLSMO}, 
as shown in Fig.\ \ref{Ndep}(b), indicates an abrupt opening of 
the energy gap below the critical thickness. 
This means that the paramagnetic component already exists slightly above the 
critical thickness of the MIT and the ferromagnetic component persists below it 
as an ferromagnetic insulating (FM-I) phase. 
Therefore, the present results suggest that a paramagnetic metallic (PM-M) 
or paramagnetic insulating (PM-I) phase starts to appear in the 
ferromagnetic metallic (FM-M) phase from slightly above the critical 
thickness, and that 
the FM-I and the PM-I phases coexist below the critical thickness of MIT. 

In order to obtain the information about changes in the electronic structure 
across the MIT, we examine the thickness dependence of the line shapes and the 
energy positions of the XAS and XMCD spectra. 
Comparing the experimental XAS spectra with those of 
LaMnO$_3$ (Mn$^{3+}$) \cite{Mn23XAS} and 
SrMnO$_3$ (Mn$^{4+}$) \cite{SMOXAS} [Fig.\ \ref{XASall}(b)], 
the intensities of structures a and c, which originate from Mn$^{3+}$, 
become stronger when the LSMO thickness is reduced. 
In addition, as shown in Figs.\ \ref{XASall}(b) and \ref{Ndep}(b), 
the peak positions of the Mn L$_3$ and L$_2$ edges are 
abruptly shifted to lower energies by $\sim 0.2\ \text{eV}$ between 
8 ML and 6 ML, where the thickness-dependent MIT occurs \cite{YoshiLSMO}. 
Similar peak shifts are also observed in the XMCD spectra, 
as shown in Fig.\ \ref{XASall}(c).
In the reference XAS spectra in Fig.\ \ref{XASall}(b), 
both the L$_3$ and L$_2$ edges are located at lower photon energies for 
Mn$^{3+}$ than for Mn$^{4+}$. 
From these spectral changes, we conclude that the effective hole concentration 
decreases as the LSMO thickness decreases, 
and that it suddenly drops at the critical thickness of MIT. 
Considering that bulk LSMO enters the FM-I phase in the low hole concentration 
region $0.09 \lesssim x \lesssim 0.16$ \cite{Urushi}, 
the observed valence shift towards Mn$^{3+}$ in the thin films is certainly 
related with the FM-M to FM-I transition with decreasing thickness. 

The observed valence change towards Mn$^{3+}$ with decreasing 
film thickness may be partly explained by 
the presence of the LSTO layer in the cap. 
As mentioned in Sec.\ \ref{Exp} a 1-ML-thick LSTO layer is inserted between 
the LSMO film and the STO cap layer in our samples [Fig.\ \ref{XASall}(a)]. 
The (LSO)$^{0.6+}$ layer in LSTO acts as an electron donor. 
Because the work function of STO is smaller than that of LSMO, 
the electrons supplied by the LSO layer are doped into the LSMO side 
rather than the STO side \cite{Kumi_Ti4+}. 
Thus the average Mn valence is shifted towards the $3+$ side from the 
nominal valence $3.4+$. 
This effect is more significant in the thinner films because 
the number of the doped electrons per monolayer is larger. 
However, the amount of electron charges is not enough to explain 
the observed valence change. 
There is also a possibility that some oxygen vacancies 
may exist in the LSMO films and/or STO substrates, which leads to 
additional electron doping. 
We note that such a valence change towards Mn$^{3+}$ with decreasing 
film thickness has also been reported in previous studies 
\cite{JSLee_reverse, FelipXRD}. 

The coincidence of the MIT and the abrupt valence change at 6-8 ML 
implies some connection between the MIT and the valence change. 
One possible scenario is that 
the MIT induces changes in the charge distribution in the LSMO film, 
leading to the apparent decrease of the hole concentration 
over the entire LSMO film. 
When the film is thick and metallic, the free electric charges (holes) 
will be distributed at the top and bottom interfaces of LSMO, 
so that there is no potential gradient inside the film. 
When the film becomes thinner and insulating, the holes are distributed 
over the entire LSMO layer. 
If the holes at the bottom interface are distributed 
in a more extended region, 
it may cause the abrupt Mn valence change observed by XMCD 
(the probing depth of which is 3-5 nm \cite{TEYdepth} or 8-12 ML). 
In order to clarify the relationship between the observed valence change 
and the MIT, further experiments would be required, 
especially on the depth profiles of the electronic states and 
magnetism in the LSMO thin films. 

\section{\label{Summ}Summary}
We have performed XAS and XMCD studies of LSMO thin films with 
varying thickness in order to investigate the origin of the 
thickness-dependent MIT and the concomitant loss of ferromagnetism. 
With decreasing film thickness, a gradual decrease of the ferromagnetic 
component and an increase of the paramagnetic component were observed. 
The experimental paramagnetic susceptibility was larger than the Curie law, 
indicating that spin correlations between Mn atoms are ferromagnetic. 
The Mn valence was found to approach $\text{Mn}^{3+}$ 
below the critical thickness of the MIT. 
The ferromagnetic-to-paramagnetic transition occurred gradually 
as a function of thickness, whereas the MIT and the valence change towards 
$\text{Mn}^{3+}$ took place abruptly. 
These results can be understood within the picture of mixed phases: 
the films above the critical thickness 
as a mixture of the FM-M and PM-I or PM-M phases 
while the films below the critical thickness 
as a mixture of the FM-I and PM-I phases. 
The mechanism of the valence change has to be investigated 
as a future theoretical problem. 

\begin{acknowledgments}
This work was supported by a Grant-in-Aid for Scientific Research 
from the JSPS (No. S22224005) and 
the Quantum Beam Technology Development Program from the JST. 
The experiment was done under the approval of the 
Photon Factory Program Advisory Committee 
(Proposal No. 2010G187 and No. 2010S2-001) 
and under the Shared Use Program of JAEA Facilities 
(Proposal No. 2011A3840/BL23SU). 
G.S. acknowledges support from 
Advanced Leading Graduate Course for Photon Science (ALPS) 
at the University of Tokyo and the JSPS Research Fellowships 
for Young Scientists (Project No. 26.11615). 
\end{acknowledgments}

\bibliography{ref.bib,MyNotes.bib}
\end{document}